\documentclass{raa_twocolumn}            

\usepackage{graphicx,times,amsmath,amssymb}             

\begin{document}

   \title{Circular polarization of fast radio bursts in the curvature radiation scenario
}

   \volnopage{Vol.0 (20xx) No.0, 000--000}      
   \setcounter{page}{1}          

   \author{H. Tong \and H. G. Wang}

   \institute{School of Physics and Materials Science, Guangzhou University, Guangzhou 510006, China;
   {\it tonghao@gzhu.edu.cn}\\
   }

   \date{Received~~2009 month day; accepted~~2009~~month day}

\abstract{The curvature radiation is applied to the explain the circular polarization of FRBs. Significant circular polarization is reported in both apparently non-repeating and repeating FRBs. Curvature radiation can produce significant circular polarization at the wing of the radiation beam. In the curvature radiation scenario, in order to see significant circular polarization in FRBs (1) more energetic bursts, (2) burst with electrons having higher Lorentz factor, (3) a slowly rotating neutron star at the centre are required. Different rotational period of the central neutron star may explain why some FRBs have high circular polarization, while others don't. Considering possible difference in refractive index for the parallel and perpendicular component of electric field, the position angle may change rapidly over the narrow pulse window of the radiation beam. The position angle swing in FRBs may also be explained by this non-geometric origin, besides that of the rotating vector model.
\keywords{fast radio bursts -- stars: magnetar -- pulsars: general -- polarization}
}

   \authorrunning{Tong \& Wang}            
   \titlerunning{Circular polarization of FRBs}  

   \maketitle

%
%
\section{Introduction}           

Fast radio bursts (FRBs) are short duration radio bursts (Lorimer et al. 2007; Thortron et al. 2013), whose origin is unclear at present (Petroff et al. 2019; Zhang 2020). FRBs may be related to pulsars and magnetars (CHIME/FRB collaboration 2020; Bochenek et al. 2020; Lin et al. 2020; Kirsten et al. 2021). At present, radio observation is the main channel via which the physics of FRBs can be studied (Petroff et al. 2019). Therefore, the full temporal, spectra, and polarization characteristics should be explored. Combining the temporal and spectra observations, interesting frequency ``down drifting'' phenomena were observed (Hessels et al. 2019; CHIME/FRB collaboration 2019a, b; Fonseca et al. 2020). The frequency down drifting may be explained similar to the radius-to-frequency mapping of pulsars (Wang et al. 2019; Tong et al. 2022).

Previously, circular polarization is reported in FRB 181112 (Cho et al. 2020, circular polarization fraction up to $34\%$), and in FRB 190611 (Day et al. 2020, circular polarization fraction of one subpulse up to $57\%$). Recently, significant circular polarization is reported in one repeating FRB 20201124A (Hilmarsson et al. 2021; Kumar et al. 2021; Xu et al. 2021). The circular polarization fraction can be as high as $75\%$ (Xu et al. 2021), which is very high even compared with that of pulsars and magnetars (Dai et al. 2021). The circular polarization may be of intrinsic origin (Hilmarsson et al. 2021), especially for the bursts with high circular polarization fraction (Xu et al. 2021).

Even in the case of pulsars, circular polarization is viewed as a complex phenomena (Han et al. 1998; You \& Han 2006; Dai et al. 2021). At the same time, our previous experiences in pulsar coherent radio emission tell us that curvature radiation may be able to produce significant circular polarization (Michel 1987; Gil \& Snakowski 1990a; Mitra et al. 2009; Gangadhara 2010; Wang et al. 2012; see Mitra 2017 for a review). Therefore, by applying the curvature radiation in neutron star magnetosphere to FRBs, we may tell under what condition the high circular polarization of FRBs can be produced\footnote{This means that we are assuming magnetospheric models like that of pulsars and magnetars for FRBs. For a review of FRBs models see Petroff et al. 2019; Zhang 2020; Lyubarsky 2021. Irrespective of the central engine, if the curvature radiation is the radiation mechanism for FRBs, the discussions here may be applied.}. Circular polarization in combination with polarization position angle swing may give us new clues about the physics of FRBs (Kumar et al. 2021; Xu et al. 2021).

Circular polarization may be caused by intrinsic or propagation effects (Han et al. 1998; You \& Han 2006; Wang et al. 2021; Beniamini et al. 2022). Here the intrinsic curvature radiation process is considered. Plasma effect may also account for the circular polarization of FRBs (Melikidze et al. 2000; Gil et al. 2004; Wang et al. 2021), which is a different way of modeling compared with the curvature radiation here. The expression for the amplitude of curvature radiation can be found in Jackson (1999)'s textbook (eq.(14.77) there; see also the Appendix for discussions). Using the words of Lyubarsky (2021), ``coherent emission could be simply described by formulas from the Jackson's textbook". We follow Gil \& Snakowski (1990a) and subsequent works for the symbol system for labeling the curvature radiation. More recent works on curvature radiation in the neutron star magnetosphere can be found in Gangadhara (2010), Wang et al. (2012), Gangadhara et al. (2021) etc. The basic picture is similar.

Curvature radiation in the neutron star magnetosphere is calculated in Section 2. Under the curvature radiation scenario, the implications for FRBs are presented in Section 3. Discussion and conclusion are given in Section 4. Major steps of the deduction of curvature radiation are presented in the Appendix.

\section{Curvature radiation in the neutron star magnetosphere}

\subsection{Single particle case}

\begin{figure}
  \centering
  \includegraphics[width=0.5\textwidth]{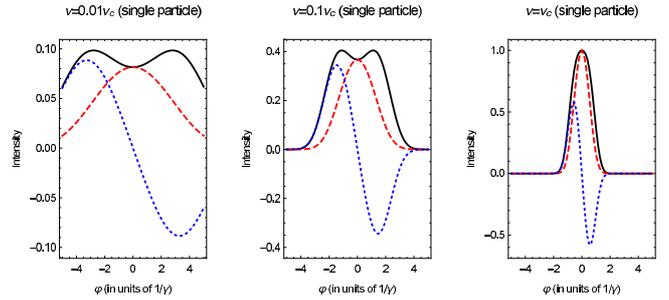}\\
  \caption{Polarization of curvature radiation as function of viewing angle for different frequency, single particle case. The black solid line is the total intensity, the red dashed line is the linear polarization intensity, the blue dotted line is the circular polarization intensity. $\nu_c$ is the typical frequency of curvature radiation. The intensity is normalized at the peak intensity at $\varphi=0$ (i.e. for $\nu = \nu_c$ and $\varphi=0$). High circular polarization is evident at the wing of radiation beam. See text for details.}
  \label{fig_intensity}
\end{figure}

\begin{figure}
  \centering
  \includegraphics[width=0.5\textwidth]{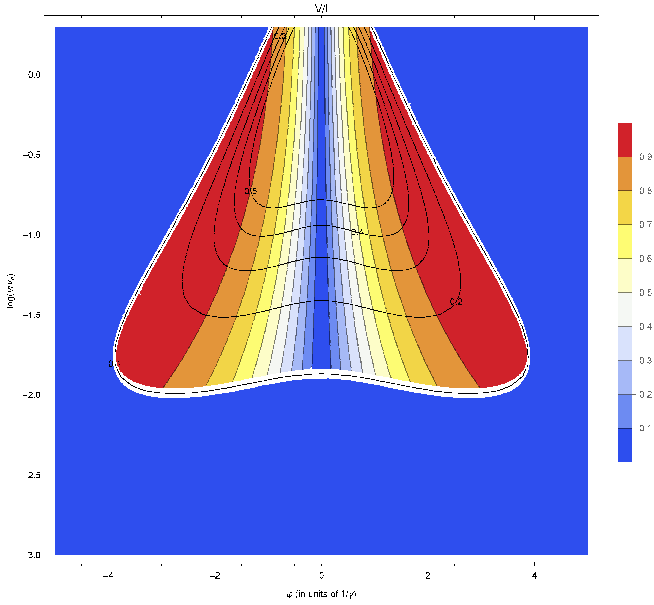}\\
  \caption{Circular polarization fraction as a function of viewing angle and frequency. The coloring shows the value of circular polarization fraction $V/I$. Overlaid black contour lines are the intensity normalized to the maximum intensity at $\varphi=0$. When the intensity is less than ten percent of the maximum intensity, $V/I$ is set to be zero. High circular polarization fraction is reached at the wings and at low frequency.}
  \label{fig_circular}
\end{figure}

The basic assumption is that highly relativistic particles flowing along curved magnetic field lines. During this process, coherent curvature radiation is emitted. This has been discussed for the radio emission of pulsars (Ruderman \& Sutherland 1975; Buschauer \& Benford 1976; Benford \& Buschauer 1977; Ochelkov \& Usov 1980; Gil \& Snakowski 1990a; Gangadhara 2010; Wang et al. 2012; Gangadhara et al. 2021) and FRBs (Kumar et al. 2017; Ghisllini \& Lacotelli 2018; Katz 2018; Yang \& Zhang 2018; Wang et al. 2022). The electric field of curvature radiation and the definition of Stokes parameters are presented in the appendix. The polarization characteristics can be seen from the four Stokes parameters. The Stokes parameters in the single particle case is shown in Figure \ref{fig_intensity} (similar to Figure 6 in Gil \& Snakowski 1990a). The typical frequency\footnote{The usually cited characteristic frequency of curvature radiation is: $\nu_c =0.24 \gamma^3 c/\rho$ (Jackson 1999; Ghisellini \& Lacotelli 2018). It is similar but differs a little with the definition here (Gil \& Snakowski 1990a).} $\nu_c = 0.2 \gamma^3 c/\rho$ corresponds to the maximum of Stoke parameter $I$ for the angle $\varphi=0$ (Gil \& Snakowski 1990a), where $\gamma$ is the electron Lorentz factor, $c$ is the speed of light in vacuum, $\rho$ is the curvature radius. From Figure \ref{fig_intensity}, it can be seen that:
\begin{enumerate}
  \item At $\nu= \nu_c$, the curvature radiation is highly linearly polarized at $\varphi=0$ (For $\varphi=0$, the observer sees the particle orbit edge on). The circular polarization fraction is low and changes sign at $\varphi=0$.
  \item At lower frequency, e.g. $\nu= 0.1\nu_c$, the intensity becomes lower and the angular spread is wider.
  \item At the wings of the emission beam, e.g. $\varphi \sim 1/\gamma$, the circular polarization intensity is high and the sign of circular polarization is constant.
\end{enumerate}
The angle $\varphi$ may be view as the pulse phase for pulsar single pulses (Gil \& Snakowski 1990a). The detailed relation can be found using spherical trigonometry in neutron star magnetospheres (Tong et al. 2021). Irrespective of the underlying magnetospheric geometry (inclination angle, viewing angle, dipolar or multipolar magnetic field etc), significant circular polarization will be present at the wing of the radiation beam. The contours of intensity as a function of viewing angle and frequency are shown as black lines in Figure \ref{fig_circular}.

For our present focus, the fraction of circular polarization $V/I$ as a function of both viewing angle and frequency is shown in Figure \ref{fig_circular}.
At the wings and low frequency, the total intensity will be lower. When the total intensity $I(\nu,\varphi)$ is less than $0.1 I(\nu_c,0)$ (i.e., ten percent of the maximum intensity), it is considered as unobservable and $V/I$ is set to zero. The criterion of ten percent the maximum intensity is arbitrary. It is only to show the effect of decreasing intensity at the wing and low frequency. From Figure \ref{fig_circular}, it can be seen that high circular polarization fraction is possible at large viewing angle and at low frequency (e.g., at $\nu \sim 0.1 \nu_c$ and $\varphi \sim 1/\gamma$).

\subsection{Additional factors: flat bunch, phase difference, and rotation of the neutron star}

The above calculation is only for the single particle case (or giant charge). In the neutron star magnetosphere, additional factor will complicate the process. We consider three of them following serial works of Gil and collaborators. The assumptions of a flat bunch, possible phase differences are rather simple minded. It is to show the possible physics inside a real neutron star magneosphere. More detailed calculations under a specified magnetospheric geometry, particle energy distribution etc can be seen in Yang \& Zhang (2018), Wang et al. (2022). After considering these effecrts, significant circular polarization may always be seem at the wing of radiation beam (Wang et al. 2021). This is consistent with the conclusions here.

The radio emission of pulsars and FRBs must be coherent. The basic coherent unit of radiating particle system may be in the form of bunches. The bunch may be generated by star quakes or magnetic reconnection etc (Wang et al. 2018; Wadiasingh \& Timokhin 2019). There must be more than one bunch in the emitting region of FRBs. If the length is larger than a half-wavelength, the emissions would become incoherent. The shape or length of a system of bunches may affect the Stokes parameters (Yang \& Zhang 2018; Wang et al. 2021, 2022). For the case of simplicity, a flat bunch is considered (Gil \& Snakowski 1990a). For a flat bunch, its radial extension is ignored. Only its distribution in the azimuthal direction is considered (Gil \& Snakowski 1990a):
\begin{equation}\label{eqn_bunch}
  \kappa(\varphi_0) \propto \exp(-\varphi_0^2/2\sigma_{\varphi}^2),
\end{equation}
where $\varphi_0$ is the phase from the centre of the bunch, $\sigma_{\varphi}=1/\gamma$ is the typical width of the bunch. The total electric field (and its Fourier transformation) of the particles in the bunch may be added coherently (Gil \& Snakowski 1990a) or non-coherently (Gil \& Snakowski 1990b). Coherent addition of electric field for the whole bunch is assumed (Gil \& Snakowski 1990a; for more physical treatment see Yang \& Zhang 2018; Wang et al. 2022):
\begin{equation}
  \hat{\mathcal E} (\nu,\varphi) = \int {\mathcal E}(\nu,\varphi-\varphi_0) \kappa(\varphi_0) d\varphi_0,
\end{equation}
which is valid for both the parallel and the perpendicular component.

If the magnetosphere or the medium has different refractive index for the parallel and perpendicular component, additional phase difference will be generated (Gil \& Snakowski 1990b; for more physical treatment see Wang et al. 2022):
\begin{equation}\label{eqn_dphi}
  {\mathcal E}_{\bot} \rightarrow e^{i\delta_{\phi}} {\mathcal E}_{\bot},
\end{equation}
where $\delta_{\phi}= 2\pi \nu \, \delta_t$ is the additional phase difference between the perpendicular and parallel component, $\delta_t= \frac{L}{c}(n_{\|} - n_{\bot})$ is the difference in propagation time difference over a region of length $L$, $n_{\|}$ and $n_{\bot}$ are the refractive index for the parallel and perpendicular component of electric field, respectively. In the neutron star magnetosphere, the phase difference $\delta_{\phi}$ may be a random large number (Gil et al. 1993).

\begin{figure}
  \centering
  \includegraphics[width=0.5\textwidth]{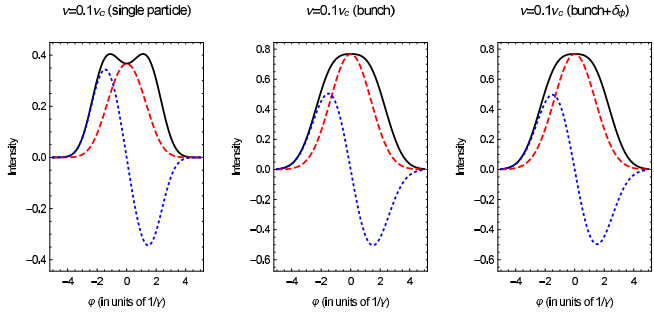}\\
  \caption{Polarization of curvature radiation, for a bunch of particles and possible phase difference. A low frequency of $\nu=0.1\nu_c$ is chosen. A phase difference of $\delta_{\phi} = 10^{\circ}$ is assumed. High circular polarization still exist at the wing of the emission beam due to the density gradient of the particle bunch.}
  \label{fig_bunch}
\end{figure}

\begin{figure}
  \centering
  \includegraphics[width=0.5\textwidth]{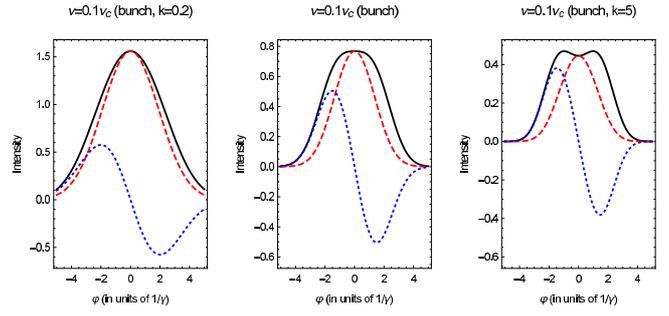}\\
  \caption{Polarization of curvature radiation, for different density of particle bunch. The middle panel corresponds to the default value of $k=1$. For higher $k$ value, the density distribution of the particle bunch is steeper. High circular polarization still exist at the wing of pulse.}
  \label{fig_density}
\end{figure}

The corresponding intensity considering a bunch of particles and possible phase difference is shown in Figure \ref{fig_bunch}. Significant circular polarization still exist due to the density gradient of the particle bunch (eq.(\ref{eqn_bunch})). A general form of particle distribution in the bunch is:
\begin{equation}
\kappa(\varphi_0) \propto \exp(-k\varphi_0^2/2\sigma_{\varphi}^2).
\end{equation}
A default value of $k=1$ is assumed. For a comparison of different particle density $k=0.2$, $k=1$, $k=5$, the polarization characteristic is shown in Figure \ref{fig_density}. Higher value of $k$ corresponds to steeper density distribution. Irrespective of the density distribution, high circular polarization still exist at the wing of the pulse profile.

\begin{figure}
  \centering
  \includegraphics[width=0.5\textwidth]{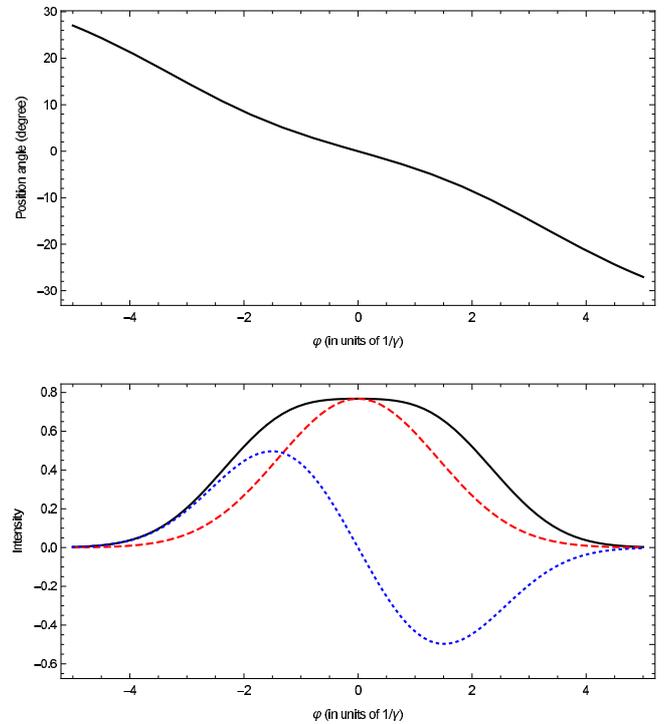}\\
  \caption{Polarization intensity and position angle as a function of viewing angle. A phase difference of $\delta_{\phi} = 10^{\circ}$ is assumed. The position angle changes rapidly over the narrow pulse window of the radiation beam. The calculation can be compared with the observations: panel B-2 and B-3 in Figure 3 in Kumar et al. (2021), Burst 1472 in Figure 2 in Xu et al. (2021).}
  \label{fig_PA}
\end{figure}

The inclusion of a possible phase difference will not change the intensity significantly. The most import aspect of a phase difference $\delta_{\phi}$ is the change in the position angle $\psi$. Without the phase difference, the position angle is constant at difference phases (Gil \& Snakowski 1990a). When the phase difference is finite, the position angle will change rapidly across the narrow phase of the single pulse (Gil \& Snakowski 1990b). For a phase difference of $\delta_{\phi}=10^{\circ}$, the result is shown in Figure \ref{fig_PA}.

In the case of pulsars, due to the rotation of the neutron star, different bunches at difference phase will be added in-coherently. This will result in cancelation of circular polarization. The final result will be no or low circular polarization in the centre of the pulse, low circular polarization with constant sign at the wing of the pulse (Gil et al. 1995).
In order to see significant circular polarization in pulsars or FRBs, slow or non-rotating magnetosphere are required\footnote{For pulsar integrated pulse, the requirement is non-drifting of subpulse (Gil et al. 1995). Here we focus on single pulse or FRBs.}. According to our previous discussions, for circular polarization, low frequency are preferred, e.g. $\nu = 0.1 \nu_c = 0.02 \gamma^3 c/\rho$. Therefore, the typical Lorentz factor is:
\begin{equation}
  \gamma = 150 \nu_9^{1/3} \rho_6^{1/3},
\end{equation}
where $\nu_9$ is the radio frequency in units of $10^9 \ \rm Hz$, $\rho_6$ is the curvature radius in units of $10^6 \ \rm cm$ (which may be valid for radio emission near the neutron star magnetosphere, Rankin 1993). For a slowly rotating neutron star magnetosphere, the rotational phase in time $\Delta t$ should be smaller than the typical bunch size:
\begin{equation}
  \Omega \Delta t \ll \frac{1}{\gamma},
\end{equation}
where $\Omega$ is the neutron star rotational angular velocity. The constraint on the neutron star rotational period is:
\begin{equation}\label{eqn_period}
  P \gg 1 \ \nu_9^{1/3} \rho_6^{1/3} \Delta t_{\rm ms} \ \rm s,
\end{equation}
where $\Delta t_{\rm ms}$ is the typical pulse width in units of milliseconds. For single pulse of pulsars or FRBs, the pulse width may be several milliseconds. Then the neutron star rotational period should be larger than several seconds in the case of slow rotation. Normal magnetar will rotational periods of several seconds (Kaspi \& Beloborodov 2017) may fulfill this requirement. Young magnetars with periods of several milliseconds can not be taken as slow rotators. Therefore, the period of the central neutron star may determine whether the circular polarization of FRBs is low or high.

\section{Implications for FRB circular polarization}

\begin{figure}
  \centering
  \includegraphics[width=0.45\textwidth]{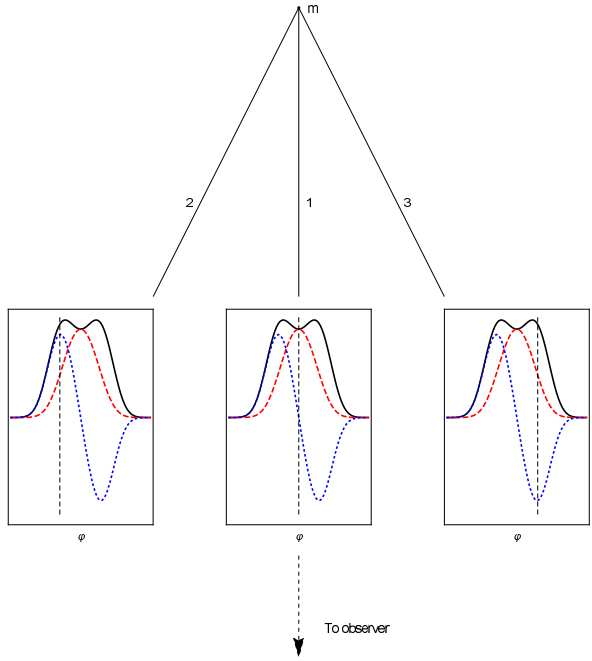}\\
  \caption{Schematic figure for circular polarization in FRBs. $m$ is the magnetic pole. The particle orbit is along magnetic field line 1. Significant circular polarization is present if the observer views the particle in the direction along line 2 or line 3.}
  \label{fig_FRB}
\end{figure}

The recently reported circular polarization in the repeating FRB 20201124A can reach a fraction up to $75\%$ (Hilmarsson et al. 2021; Kumar et al. 2021; Xu et al. 2021). According to the calculations in the previous section, a general picture for the circular polarization in FRBs is shown in Figure \ref{fig_FRB}. In Figure \ref{fig_FRB}, the particle moves along magnetic field line 1. If it is viewed edge on ($\varphi=0$), the observer will see a low level of circular polarization. If it is viewed at an angle $\varphi \sim 1/\gamma$ (line 2 or line 3), then significant circular polarization is expected and the sign of circular polarization should be constant. The differences between line 2 and line 3 is  that the sign of their circular polarization is opposite.

More quantitatively, in order to see significant circular polarization in FRBs, it is required that:
\begin{enumerate}
  \item The radiation beam is viewed at the wing. At the centre of the beam, no significant circular polarization is expected. At the wing of the beam, the intensity will be lower (see Figure \ref{fig_intensity} and \ref{fig_circular}). Therefore, more energetic burst is preferred if it is viewed at the wing.
  \item The curvature radiation is observed at the low frequency. At the peak frequency $\nu_c$, the emission beam is very narrow. Only at lower frequency, e.g. $\nu=0.1\nu_c$, the emission beam is wider. Only at lower frequency, it is more probable to view the emission beam at the wing (see Figure \ref{fig_circular}). Therefore, if the electrons in the FRB have higher Lorentz factor, it is more probable that the observational frequency lies at the low frequency end of curvature radiation.
  \item The magnetosphere is non-rotating or slowly rotating. In-coherent addition of different particle bunch at difference phase will result in cancelation of circular polarization (Lyne \& Graham-Smith 2012; Gil et al. 1995). Therefore, a slowly rotating neutron star is preferred in order to see significant circular polarization (see eq.(\ref{eqn_period})).
      If the neutron star inside FRB 121102 has a small rotational period, and if the neutron star inside FRB 20201124A has a large rotational period, it may explain why these two FRBs have different circular polarization characteristics (low for FRB 121102, Michilli et al. 2018; high for FRB 20201124A, Hilmarsson et al. 2021, Kumar et al. 2021, Xu et al. 2021).
      This may also explain why magnetars have higher circular polarization than normal pulsars (Dai et al. 2021, Figure 1 there).
\end{enumerate}

Considering possible phase difference between the perpendicular and parallel component, the position angle may vary rapidly over the narrow pulse of the radiation beam (see Figure \ref{fig_PA}). This is dubbed as non-geometric swing of position angle (Jil \& Snakowski 1990b). For the position angle swing not associated with circular polarization (Luo et al. 2020) and associated with significant circular polarization (Kumar et al. 2021; Xu et al. 2021), there are at least three possible explanations:
\begin{enumerate}
  \item The position angle swing is due to a rotating magnetosphere (i.e., rotating vector model, Radhrishan \& Cooke 1969). In the rotating vector model, a monotonic variation of position is expected for both pulsars and magnetars (Tong et al. 2021).
  \item The position angle swing is due to non-geometric swing (Jil \& Snakowski 1990b). The neutron star magnetosphere is rotating. But the rapid swing over the narrow pulse window is due to non-geometric origin.
  \item It can not be excluded that the neutron star magnetosphere is not rotating. If the parallel and perpendicular components of the electric field of curvature radiation have a varying refractive index during their propagation, the observer will also see a varying position angle similar to Figure \ref{fig_PA}. For a non-rotating magnetosphere, if $n_{\|}- n_{\bot}$ does not vary with time, a constant position angle is expected. If $n_{\|}- n_{\bot}$ vary with time monotonously, a monotonic variation of position angle is expected. If $n_{\|}- n_{\bot}$ vary with time non-monotonously, then a diverse swing of position angle is expected.
\end{enumerate}
Both points (2) and (3) are due to non-geometric origin for variation of position angle.
The integrated pulse of pulsars may follow the rotating vector model. However, the position angle of single pulse may differ significantly with the integrated pulse (Gil 1992; Gil et al. 1997). Considering the experiences in pulsar single pulses, the non-geometric origin (e.g., point 2) may be preferred for the rapid and diverse position angle swing in FRBs (Luo et al. 2020; Kumar et al. 2021; Xu et al. 2021).

In point (2), for a slowly rotating neutron star magnetosphere, during the burst duration, the observers may only see part of the radiation beam. The right part (i.e. $\varphi >0$) of Figure \ref{fig_PA} shows a decreasing position angle accompanied by a significant circular polarization. This can be compared with the observations: panel B-2 and B-3 in Figure 3 in Kumar et al. (2021), Burst 1472 in Figure 2 in Xu et al. (2021). To be more specific, for burst P05 (Kumar et al. 2021) and burst 1472 (Xu et al. 2021) from FRB 20201124A, we are seeing the burst from the wing of the radiation beam, the observational frequency lies at the low frequency end of the curvature radiation, the central neutron star may be slowly rotating, during the burst duration only part of the radiation beam is seen. The net result is a high circular polarization accompanied with a swinging position angle. Compared with with Burst 1472, Burst 1112 has a different sign of circular polarization and a different slope of position angle (Xu et al. 2021, Figure 2 there). This may be caused by a different phase difference $\delta_{\phi}$ between the perpendicular and parallel component of the electric field (eq.({\ref{eqn_dphi})). While the central part (around $\varphi=0$) of  Figure \ref{fig_PA} may corresponds to the varying position angle with low circular polarization (Luo et al. 2020, Figure 1 there). As discussed above, a small rotational period of the central neutron star may also result in low circular polarization.

\section{Discussion and conclusion}

The curvature radiation has been studied in the case of pulsars. The polarization characteristics of curvature radiation in pulsars have been investigated in various aspects (Michel 1987; Gil \& Snakowski 1990a; Gandhara 2010; Wang et al. 2012). The curvature radiation has also been proposed as a possible radiation mechanism for FRBs, focusing on the luminosity and spectra etc. Polarization of single pulses in pulsars indicates that the radio emission is the X-mode of curvature radiation (Mitra et al. 2009). Using similar arguments, Lu et al. (2019) proposed that the FRB radio emission is also the X-mode of curvature radiation.

We mainly consider curvature radiation in vacuum here. FRB radio emission may be choked by the neutron star magnetospheric plasma (Beloborodov 2021). However, considering that the FRB emission is like a strong electromagnetic pulse. The radiation pressure of FRB emission may dominate over the gravitational pull and push the plasma ahead of it and result in a near vacuum condition (Section 2.2 in Wang et al. 2022). Even considering possible plasma effects, the change may be mainly quantitative (Qu \& Zhang 2022).

From previous experiences in pulsars and FRBs, there are three ways to generate circular polarization.
\begin{enumerate}
  \item By the radiation mechanism, e.g. curvature radiation. Both synchrotron and curvature radiation can generate significant circular polarization (Michel 1987; Gil \& Snakowski 1990a). However, inverse Compton scattering from one giant charge can only result in linear polarization (Xu et al. 2000; Zhang 2022). Therefore, if FRB is generated by a giant charge (one particle bunch), curvature radiation may explain the significant circular polarizations, while inverse Compton scattering can not.
  \item By coherent superposition of electric field from a system of bunches. From knowledge in optics, coherent addition of two perpendicular electric field will result in circular polarization intensity. Significant circular polarization may be generated in this way, while the electric field can be from both curvature (Wang et al. 2022) and inverse Compton scattering process (Xu et al. 2000).
  \item By propagation and/or plasma effect. The above two points can occur in vacuum or near vacuum case. The magnetospheric plasma may also account for the circular polarization of pulsar and FRB radio emissions (cyclotron absorption and/or propagation, Petrova \& Lyubarskii 2000; Melrose 2003; Melrose \& Luo 2004; Wang et al. 2010; Kumar et al. 2022)
\end{enumerate}
From the above three points, the difference between our calculations and Wang et al. (2022) is: we mainly consider curvature radiation from a giant charge and discuss its various consequences in the case of FRBs. Wang et al. (2022) also include the effect of coherent superposition of electric field from a system of bunches.

The polarization characteristics of curvature radiation (e.g., circular polarization discussed here) applies to both the integrated and single pulse of pulsars and FRBs, if the curvature radiation is the radiation mechanism for pulsars and FRBs. This applies to both repeating FRBs and apparently non-repeating ones, if curvature radiation is at work in both cases. In this respect, FRBs are the analogy of pulsar single pulses.

In order to understand the circular polarizations in FRBs (Cho et al. 2020; Day et al. 2020; Hilmarsson et al. 2021; Kumar et al. 2021; Xu et al. 2021) in the curvature radiation scenario, we conclude that:
\begin{enumerate}
  \item The curvature radiation can naturally produce significant circular polarization at the wing of the radiation beam. This have already been known in the case of pulsars (Michel 1987; Gil \& Snakowski 1990a).
  \item In order to see significant circular polarization in FRBs, (1) more energetic burst, (2) burst with electrons having higher Lorentz factor, (3) slow rotators inside the burst are preferred.
  \item The position angle swing in FRBs may be produced by either the rotating vector model or the non-geometric swing. Non-geometric origin may be preferred for the rapid and diverse position angle swing in FRBs.
\end{enumerate}

\section*{Acknowledgments}
This work is supported by National SKA Program of China (No. 2020SKA0120300) and NSFC (12133004). We'd like to thank the referee for helpful suggestions which improved the quality of this paper.

\appendix

\section{Electric field of curvature radiation}

Michel (1987) first noted that synchrotron radiation may produce significant circular polarization and a sign change (i.e. sense reversal) in pulsars. Gil \& Snakowski (1990a) calculated the electric field of curvature radiation. The amplitude of curvature can also be found in Jackson's textbook. Major steps are listed here in order to make clear the physics involved.

Following Gil \& Snakowski (1990a), the radiation field of a moving charge is (Jackson 1999):
\begin{equation}
  {\bf E}(t) = \frac{e}{c R_0(\tau)}%
  \frac{{\bf n} \times [ ({\bf n} - {\boldsymbol \beta}) \times \dot{\boldsymbol \beta} ]}{(1 - {\boldsymbol \beta} \cdot {\bf n})^3},
\end{equation}
where $R_0(\tau)$ is the distance from the source point to the field point at the retarded time $\tau$, ${\boldsymbol \beta}$ is the dimensionless particle velocity, ${\bf n}$ is the unit vector along the wave propagation direction ${\bf R}_0$, approximated as the unit vector along ${\bf R}$, see Figure \ref{fig_cur} (or Figure 14.9 in Jackson 1999). Fourier transformation of the radiation field is
\begin{eqnarray}
  {\mathcal E} &=& \int_{-\infty}^{\infty} {\bf E}(t) \exp(i 2\pi \nu t) dt \\\nonumber
  \label{eqn_magnetic_field}
  &=& \frac{e}{c R} \exp(i 2\pi \nu R/c) \int_{-\infty}^{\infty}%
  \frac{{\bf n} \times [ ({\bf n} - {\boldsymbol \beta}) \times \dot{\boldsymbol \beta} ]}{(1 - {\boldsymbol \beta} \cdot {\bf n})^2}\\
  && \times \exp(i 2\pi \nu (\tau - {\boldsymbol \rho} \cdot {\bf n}/c)) d\tau,
\end{eqnarray}
where the integration is transformed to the retarded time $\tau$. In the present case, the following three quantities are all small numbers: $\theta$, $\varphi$, and $\delta=1/\gamma$. Under these approximations, the integrand can be approximated, especially:
\begin{equation}
  {\bf n} \times [ ({\bf n} - {\boldsymbol \beta}) \times \dot{\boldsymbol \beta} ] = \pi \nu_0 (\delta^2+\varphi^2-\theta^2, 2\theta \varphi,0),
\end{equation}
where $\nu_0=\beta c/(2\pi \rho)$ is the fundamental frequency of curvature radiation.
Therefore, the electric field (which is its Fourier transformation) of the curvature radiation is:
\begin{equation}
  {\mathcal E}= \mathcal{E}_{\|} \hat{x} + \mathcal{E}_{\bot} \hat{y},
\end{equation}
where the parallel and perpendicular component are respectively:
\begin{eqnarray}
\nonumber
  \mathcal{E}_{\|} &=& \frac{2e}{R c} \exp(i 2\pi \nu R/c) \int_{-\infty}^{\infty} \frac{\delta^2 + \varphi^2 -\theta^2}{(\delta^2 + \varphi^2 +\theta^2)^2}\\
  &&\times \exp\left\{ \frac{i\nu}{2\nu_0} \left[ (\delta^2 + \varphi^2) \theta + \frac{\theta^3}{3}\right] \right\} d\theta \\
\nonumber
    \mathcal{E}_{\bot} &=& \frac{2e}{R c} \exp(i 2\pi \nu R/c) \int_{-\infty}^{\infty} \frac{2\theta \varphi}{(\delta^2 + \varphi^2 +\theta^2)^2}\\
    &&\times \exp\left\{ \frac{i\nu}{2\nu_0} \left[ (\delta^2 + \varphi^2) \theta + \frac{\theta^3}{3}\right] \right\} d\theta.
\end{eqnarray}
Using the two formulae (Pacholczyk 1970)
\begin{eqnarray}
  \nonumber
  &&\int_{-\infty}^{\infty} \exp [i s (w^2 u + \frac13 u^3)] \frac{w^2-u^2}{(w^2+u^2)^2} du\\
   &&= \frac{2}{\sqrt{3}} s w^2 K_{2/3}(\frac23 s w^3) \\
   \nonumber
  &&\int_{-\infty}^{\infty} \exp [i s (w^2 u + \frac13 u^3)] \frac{2u}{(w^2+u^2)^2} du \\
  &&= i\frac{2}{\sqrt{3}} s w K_{1/3}(\frac23 s w^3),
  \end{eqnarray}
where $K_{2/3}$ and $K_{1/3}$ are modified Bessel functions, the electric field of curvature radiation can be rewritten as
\begin{eqnarray}
\nonumber
  {\mathcal E}_{\|} &=& \frac{2e}{\sqrt{3} R c \nu_0} e^{i 2\pi \nu R/c}\ \nu (\delta^2 + \varphi^2)\\
  \label{eqn_A10}
  &&K_{2/3}\left[ \frac{\nu}{3\nu_0}(\delta^2 + \varphi^2)^{3/2} \right] \\
  \nonumber
  {\mathcal E}_{\bot} &=& i \frac{2e}{\sqrt{3} R c \nu_0} e^{i 2\pi \nu R/c}\ \nu \varphi (\delta^2 + \varphi^2)^{1/2}\\
  \label{eqn_A11}
  &&\times K_{1/3}\left[ \frac{\nu}{3\nu_0}(\delta^2 + \varphi^2)^{3/2} \right].
\end{eqnarray}
The pure imaginary number for the perpendicular component means that it is out of phase by $\pi/2$ with the parallel component. This is a necessary condition for the presence of circular polarization.

During the calculations, the common phase factor of $e^{i 2\pi \nu R/c}$ can be dropped. The frequency $\nu$ can be expressed in units of $\nu_c$. The viewing angle $\varphi$ can be expressed in units of $\delta=1/\gamma$. Defining
$\mathcal E_0 = \frac{2e}{\sqrt{3} R c \nu_0}$, expressing ${\mathcal E}_{\|}$ and ${\mathcal E}_{\bot}$ in units of $\mathcal E_0 \nu_c \delta^2$, the final expression can be simplified:
\begin{eqnarray}
  {\mathcal E}_{\|} &=& \nu (1+\varphi^2) K_{2/3}\left[ 0.42\nu (1+\varphi^2)^{3/2} \right] \\
  {\mathcal E}_{\bot} &=& i \nu \varphi (1+\varphi^2)^{1/2} K_{1/3}\left[ 0.42\nu (1+\varphi^2)^{3/2} \right].
\end{eqnarray}
The final intensity (total, linear, and circular) will all be normalized by the maximum intensity when $\varphi=0$. That is the intensity $I=1$ corresponds to the maximum intensity at phase $\varphi=0$ (when the particle orbit is viewed edge on).

The integration in equation (\ref{eqn_magnetic_field}) may be simplified by integration by parts (using eq.(14.66) in Jackson 1999).
The above deduction process may be simplified in this way.

If following the treatment of Jackson (1999), using the coordinate system in Figure 14.9 and the integration formula of eq.(14.78) in Jackson (1999), the final result is the same as eq.(\ref{eqn_A10}) and eq.(\ref{eqn_A11}), except for a common factor of $-1$ for the parallel and perpendicular component. This is due to a different orientation of the coordinate axis in Figure 14.9 in Jackson (1999) and Figure \ref{fig_cur} here. A common phase factor can be dropped as discussed above.

Furthermore, in both Figure 14.9 in Jackson (1999) and Figure \ref{fig_cur} here, the coordinate axis satisfies: $\hat{e}_{\bot} = \hat{n} \times \hat{e}_{\|}$. However, in
Figure 1 in Gil \& Snakowski (1990a), the coordinate axis are: $\hat{e}_{\bot} = -\hat{n} \times \hat{e}_{\|}$. Therefore, following the procedure in Gil \& Snakowski (1990a), the final result (eq.(12) in Gil \& Snakowski 1990a) would contain a minus sign for the perpendicular component. The conclusions there are unaffected by this typo.

In summary, the Fourier transformation of electric field of curvature radiation is expressed in eq.(\ref{eqn_A10}) and eq.(\ref{eqn_A11}). This is our starting point. We also follow the IAU/IEEE convention for the definition of Stokes parameters (van Straten et al. 2010; Rybicki \& Lightman 1979):
\begin{eqnarray}
  I &=& E_{\|}E_{\|}^{\ast} + E_{\bot}E_{\bot}^{\ast} \\
  Q &=& E_{\|}E_{\|}^{\ast} - E_{\bot}E_{\bot}^{\ast} \\
  U &=& 2{\mathrm {Re}} (E_{\|} E_{\bot}^{\ast}) \\
  V &=& 2{\mathrm {Im}} (E_{\|} E_{\bot}^{\ast}),
\end{eqnarray}
where $E_{\|}$ and $E_{\bot}$ are the parallel and perpendicular components of the electric field or its Fourier transformation.
The full information of radiation is characterised by the four parameters: (1) the total intensity $I$, (2) the linear polarization intensity $L=\sqrt{Q^2 + U^2}$, (3) the position angle $\psi = \frac12 \arctan (U/Q)$, and (4) the circular polarization intensity $V$ (Gil \& Snakowski 1990a).

In the case of pulsar studies, the definition of the sign of circular polarization and the sign of position angle may differ by a minus sign (Everett \& Weisberg 2001; van Straten et al. 2010). However, the physics is unchanged by these ambiguities: the presence of circular polarization, a change sign of circular polarization, and a swinging position angle etc.

\begin{figure}
  \centering
  \includegraphics[width=0.55\textwidth]{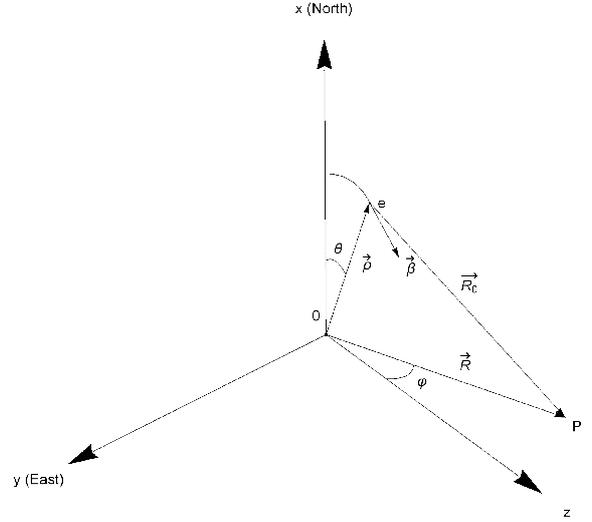}\\
  \caption{Geometry of curvature radiation. The orientation of the axis follow that of the IAU/IEEE convention (van Straten et al. 2010, figure 1 there). The particle orbit is in the x-z plane. The observer sees the curvature radiation at an angle $\varphi$ with respect to the z-axis.}
  \label{fig_cur}
\end{figure}

\label{lastpage}

\end{document}